\documentclass[a4paper,12pt]{article}
\usepackage{geometry}
\usepackage[utf8]{inputenc}
\usepackage[T1]{fontenc}
\pdfoutput=1
\usepackage{amsmath}
\usepackage{amssymb}
\usepackage{amsfonts}
\usepackage{lmodern}
\usepackage{tikz}
\usepackage{pgfplots}
\usepackage{rotating}
\usepackage{xcolor}
\usepackage{graphicx}
\usepackage{tgheros}
\usepackage{helvet}
\usepackage{amssymb}
\usepackage{tikz}
\usepackage{csquotes}
\usepackage[english]{babel}
\usepackage[]{biblatex}
\usepackage{wrapfig}
\usepackage{float}
\usepackage{multicol}
\usepackage{subcaption}
\usepackage{booktabs}
\usepackage{setspace}
\usepackage{blindtext}
\usepackage[margin=1cm]{caption}
\usepackage{longtable}
\usepackage{layouts}
\definecolor{blue}{RGB}{23,67,106}
\definecolor{red}{RGB}{186,65,12}
\definecolor{almond}{rgb}{0.94, 0.87, 0.8}

\usetikzlibrary{matrix,positioning,calc}
\usetikzlibrary{backgrounds}
\usetikzlibrary{intersections}
\usetikzlibrary{calc}
\usetikzlibrary{arrows}
\usetikzlibrary{backgrounds}
\usetikzlibrary{intersections}
\usetikzlibrary{matrix,positioning,calc}
\usetikzlibrary{backgrounds}
\usetikzlibrary{intersections}
\usetikzlibrary{calc}
\usetikzlibrary{arrows}
\usetikzlibrary{shadows}
\usetikzlibrary{backgrounds}
\usetikzlibrary{intersections}
\usetikzlibrary{arrows}

\usepackage{multicol}
\usepackage{subcaption}
\usepackage{booktabs}
\usepackage{setspace}
\usepackage{blindtext}
\usepackage[margin=1cm]{caption}
\usepackage{longtable}
\usepackage{layouts}
\definecolor{blue}{RGB}{15,47,188}
\definecolor{blue2}{RGB}{15,47,222}
\definecolor{red}{RGB}{186,65,12}
\definecolor{almond}{rgb}{0.94, 0.87, 0.8}

\title{Including the asymmetry of the Lorenz curve into measures of economic inequality}
\author{Mario Schlemmer}
\date{}
\begin{document}
\maketitle
\thispagestyle{empty}
	
\begin{abstract} 
The Gini index signals only the dispersion of the distribution and is not very sensitive to income differences at the tails of the distribution. The widely used index of inequality can be adjusted to also measure distributional asymmetry by attaching weights to the distances between the Lorenz curve and the 45-degree line. The measure is equivalent to the Gini if the distribution is symmetric. The alternative measure of inequality inherits good properties from the Gini but is more sensitive to changes in the extremes of the income distribution. 
\end{abstract} 
\noindent\textbf{Keywords:} Lorenz curve, Gini-coefficient, skewness, SAG\\
\textbf{JEL-classication:} D31, D63

\section*{Introduction}
The Gini coefficient is widely used for measuring income and wealth inequality. It can be represented geometrically by the Lorenz curve and has other desirable properties like population and scale invariance. But it is not particularly sensitive to income differences that relate to values further away from the middle of the distribution (Jenkins, 2009). Distributional asymmetry is regularly driven by remarkably low or high incomes at the tails of the distribution, and several authors have suggested the use of indices that are sensitive to the asymmetry of the distribution(Bowden, 2016; Clementi et al., 2019). Figure 1 depicts two Lorenz curves with opposite asymmetry.       
\begin{figure}[h] 	
\centering
\begin{tikzpicture}
			\begin{axis}[scale only axis, xlabel = {cummulative share of population}, xmin = 0, xmax = 1, ylabel = {cummulative share of ressources},  ymax = 1, ymin = 0, ytick pos=left]

				\draw[thick, red](5,90) -- (15,90);
				\draw[thick,black](15,90) node[right] { right-skewed};
				\draw[thick, blue2](5,83) -- (15,83);
				\draw[thick,black](15,83) node[right] { left-skewed};
				\draw[very thick,dashed, lightgray](5,76) -- (15,76);
				\draw[thick,black](15,76) node[right] { symmetric};
				
				\draw[thick,black](0,0) -- (100,100);
			
				Lorenz symmetric;
				D=[2,3,4,6,8,12,14,16,17,18]
				\draw[very thick,dashed, lightgray](0,0) -- (10,2);
				\draw[very thick,dashed, lightgray](10,2) -- (20,5);
				\draw[very thick,dashed, lightgray](20,5) -- (30,9);
				\draw[very thick,dashed, lightgray](30,9) -- (40,15);	
				\draw[very thick,dashed, lightgray](40,15) -- (50,23);
				\draw[very thick,dashed, lightgray](50,23) -- (60,35);
				\draw[very thick,dashed, lightgray](60,35) -- (70,49);
				\draw[very thick,dashed, lightgray](70,49) -- (80,65);
				\draw[very thick,dashed, lightgray](80,65) -- (90,82);	
				\draw[very thick,dashed, lightgray](90,82) -- (100,100);
				
				 \node (root0) at (0,0) {}; 
				\node (lp1) at (10,2) {}; 
				\node (lp2) at (20,5) {}; 
				\node (lp3) at (30,9) {}; 
				\node (lp4) at (40,15) {}; 
				\node (lp5) at (50,23) {}; 
				\node (lp6) at (60,35) {}; 
				\node (lp7) at (70,49) {}; 
				\node (lp8) at (80,65) {}; 
				\node (lp9) at (90,82) {}; 
				\node (lp10) at (100,100) {}; 
					
				\draw[thick,red](0,0) -- (10,6);
				\draw[thick,red](10,6) -- (20,12);
				\draw[thick,red](20,12) -- (30,18);
				\draw[thick,red](30,18) -- (40,24);	
				\draw[thick,red](40,24) -- (50,30);
				\draw[thick,red](50,30) -- (60,36);
				\draw[thick,red](60,36) -- (70,43);
				\draw[thick,red](70,43) -- (80,50);
				\draw[thick,red](80,50) -- (90,66);	
				\draw[thick,red](90,66) -- (100,100);	
				
				\draw[thick,blue](0,0) -- (10,1);
				\draw[thick,blue](10,1) -- (20,3);
				\draw[thick,blue](20,3) -- (30,6);
				\draw[thick,blue](30,6) -- (40,11);	
				\draw[thick,blue](40,11) -- (50,24);
				\draw[thick,blue](50,24) -- (60,38);
				\draw[thick,blue](60,38) -- (70,53);
				\draw[thick,blue](70,53) -- (80,67);
				\draw[thick,blue](80,67) -- (90,83);	
				\draw[thick,blue](90,83) -- (100,100);
				 
					\begin{scope}[on background layer]
					\fill[gray!30!white,on background layer,opacity=0.8] (root0.center) -- (lp1.center) --  (lp2.center) -- 
					(lp3.center) -- (lp4.center) -- 
					(lp5.center) -- (lp6.center) -- 
					(lp7.center) -- (lp8.center) -- 
					(lp9.center) -- (lp10.center) -- cycle;
					
				\end{scope}	
				
			\end{axis}
		
\end{tikzpicture}
\caption{Two Lorenz curves with opposite asymmetry that yield the same Gini coefficient(\text{G}=0.33) as a symmetric Lorenz curve. } 	
\end{figure}
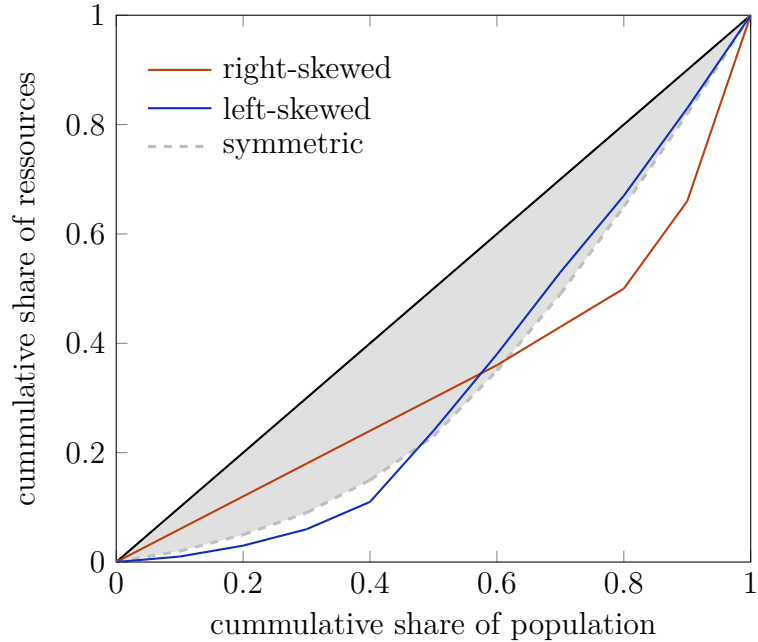
The Gini can be calculated from the distances between the Lorenz curve and the 45-degree line. We assume that we have $n$ independent observations of $x_i={x_1,x_2,x_3,...,x_n}$, say household income, the cumulative income shares are denoted $q$ and the cumulative shares of population as $p$. The Cumulative shares of households are plotted on the horizontal axis and cumulative income shares on the vertical axis. The Gini is given by:
\begin{align}
\begin{split}\\\nonumber
\text{G}=\frac{2n}{n^{2}} \sum_{i=1}^{n-1} (p_{i}-q_{i})
\end{split}
\end{align}	
\indent Subtracting the cumulative household incomes $q_i$ from $p_i$ yields the sum of distances between the cumulative distribution function and the 45-degree line. To obtain the classical Gini the sum of distances is standardized by multiplication with $2n/n^{2}$. If the distribution is symmetric the distances at opposite tails will be symmetric too,  if the distribution is not symmetric the differences can be used to calculate an extension of the Gini coefficient that is sensitive to asymmetry.
\section*{The skewness-adjusted Gini}
\indent   If the lengths of the distances in the lower and upper half of the distribution are not symmetrical this also indicates an asymmetry in the original income distribution. Such differences provide information about income inequality that goes beyond dispersion. By using a simple weighing scheme the Gini can be adjusted to take the asymmetry of the distribution into account. The weights $w$ are attached to the $n-1$ distances between the Lorenz curve and the 45-degree line. To obtain a metric that is sensitive to the remarkably high household incomes more weight is attached to distances $p_{i}-q_{i}$ as cumulative population share increases. The weights are calculated by dividing each of the first $n-1$ natural numbers through the median of the number series which is equal to \textit{n}/2. Resulting weights range from $0^{+}$ to 2, with a mean weight of 1.    
\begin{align}
\begin{split}\nonumber
\textit{$G_R$}=\frac{2n}{n^{2}} \sum_{i=1}^{n-1} (p_{i}-q_{i})w_i \text{\small{ ,where}}
\end{split}\\
\begin{split}\nonumber
w_i=\frac{2i}{\textit{n}}
\end{split} 
\end{align}	
\indent Converseley, to obtain a metric that is sensitive to remarkably low household incomes more weight is attached to the distances at the lower tail, and the weights $w^{\prime}$ decrease with population share.
\begin{align}
\begin{split}\nonumber
\textit{$G_{L}$}=\frac{2n}{n^{2}} \sum_{i=1}^{n-1} (p_{i}-q_{i})w_i^{\prime}\text{\small{ ,where}}
\end{split} \\
\begin{split}\nonumber
w_i^{\prime}=\frac{2n-2i}{n}
\end{split}
\end{align}	
\indent Because the same weights are attached in reversed order, both \textit{$G_{R}$} and \textit{$G_{L}$} are equivalent to the Gini if there is no asymmetry in the distribution. Depending on the type of skewness, one of the two metrics will be larger than $G$, and the other will be smaller than G by the same amount. The mean of the two metrics is always the Gini. A measure that is sensitive to both types of skewness is defined as:
\begin{align}\nonumber
\textit{SAG}=G+\frac{|\textit{$G_{R}$} -\textit{$G_{L}$}|}{2}
\end{align}
The skewness-adjusted Gini \textit{SAG} complements the information about statistical dispersion expressed in the Gini with information about distributional asymmetry. It takes on the value of $G_{R}$ or $G_{L}$ if the distribution is skewed. For \textit{n} large enough, the upper bounds of \textit{$G_{R}$} and {$G_{L}$} are 4/3 and 2/3 respectively, giving the \textit{SAG} an upper bound of 4/3. The standardization to a constant bound and the relationship with the Lorenz curve are useful properties. \textit{SAG} also satisfy the following three normative axioms that have been proposed to aid in identifying appropriate inequality indices:\\
	
	\indent 1. \emph{Scale invariance:} The index value does not change if all incomes are changed\par proportionally.\\
	\indent 2. \emph{Population invariance:} The index value does not change if the original popula-\par tion is replicated.\\
	\indent 3. \emph{Pigou–Dalton principle of transfers:} If a progressive transfer does not change\par the rank of the individuals in the distribution the index value decreases.\\
   
\indent From the Gini coefficient \textit{SAG} also inherits the ability to accommodate zero and negative values, and a weaker form of decomposability(Ebert, 2010). The proposed measure is related to previously described extensions of the Gini coefficient, particularly the single-series Ginis and the single-parameter Ginis(Weymark, 1981; Donaldson \& Weymark, 1980; Yitzhaki, 1983). The advantage of the \textit{SAG} over the family of single-series Ginis is that it satisfies the population principle. The single-parameter Ginis rely on the choice of an inequality-aversion parameter that expresses the concern for the share of low incomes in the distribution, but approaches of this kind can be regarded as imposing observer value judgments on the choice of metric(Bowden,2016). The proposed alternative measure of inequality \textit{SAG} avoids this problem, as only the described weights yield the popular Gini if the distribution is symmetric.

\newpage
\section*{References}
\normalsize{Bowden, R. J. (2016). Giving Gini direction: An asymmetry metric for economic\par disadvantage. \textit{Economics Letters, 138}, 96-99.}\\
\normalsize{Clementi, F., Gallegati, M., Gianmoena, L., Landini, S.,  \& Stiglitz, J. E. (2019). Mis-\par measurement of inequality: a critical reflection and new insights. \textit{Journal of Eco-} \par \textit{nomic Interaction and Coordination, 14}(4), 891-921.}\\
\normalsize{Donaldson, D., \& Weymark, J. A. (1980). A single-parameter generalization of the\par Gini indices of inequality. \textit{Journal of economic Theory, 22}(1), 67-86.}\\
\normalsize{Ebert, U. (2010). The decomposition of inequality reconsidered: Weakly decomposable\par measures.\textit{ Mathematical Social Sciences, 60}(2), 94-103.}\\
\normalsize{Jenkins, S. P. (2009). Distributionally‐sensitive inequality indices and the GB2 income\par distribution. \textit{Review of Income and Wealth, 55}(2), 392-398.}\\
\normalsize{Weymark, J. A. (1981). Generalized Gini inequality indices. \textit{Mathematical Social} \par \textit{Sciences, 1}(4), 409-430.}\\
\normalsize{Yitzhaki, S. (1983). On an extension of the Gini inequality index. \textit{International} \par \textit{ economic review}, 617-628.}
\end{document}